# Controlling the non-linear optical properties of MgO by tailoring the electronic structure

Mukhtar Hussain [1, 2] *. Hugo Pires [1]. Willem Boutu [2]. Dominik Franz [2]. Rana Nicolas [2, 3]. Tayyab Imran [4]. Hamed Merdji [2]. Marta Fajardo[1]. Gareth O. Williams[1]

[1] GoLP/Instituto de Plasmas e Fusão Nuclear-Laboratório Associado, Instituto Superior Técnico, Universidade de Lisboa, 1049-001 Lisboa, Portugal
[2] LIDYL, CEA, CNRS, Université Paris-Saclay, CEA Saclay 91191, Gif-sur-Yvette, France
[3] Lebanese American University, Natural Sciences Department, Beirut, Lebanon
[4] Group of Laser Development (GoLD)/Department of Physics, COMSATS University Islamabad, Park Road, 44000, Islamabad, Pakistan

## Abstract

The study of the non-linear response of matter to high electric fields has recently encompassed harmonic generation in solids at near-infrared (NIR) driving wavelengths. Interest has been driven by the prospect of ultrafast signal processing and all-optical mapping of electron wave-functions in solids. Engineering solid-state band structures to control the non-linear process has already been highlighted theoretically. Here, we show experimentally for the first time that second harmonic generation (SHG) can be enhanced by doping crystals of magnesium oxide (MgO) with chromium (Cr) atoms. We show that the degree of enhancement depends non-linearly on dopant concentration. The SHG efficiency is shown to increase when Cr dopants are introduced into pure MgO. A physical picture of the effect of Cr dopants is aided by density functional theory (DFT) calculations of the electronic structure for pure and doped samples. This work shows an unambiguous enhancement of the SHG efficiency by modifying the electronic structure. The observed effects are consistent with an electronic structure that facilitates the surface induced SHG and demonstrates a minimal angular dependence. This work highlights the potential of manipulating the electronic structure of solids to control or test theories of their non-linear optical response.

## 1. Introduction

Second harmonic generation (SHG) is the first perturbative non-linear response of solids at high driving fields. Efficient SHG is not possible in centrosymmetric solids due to symmetry conditions which forbid phase-matching. Near surfaces and interfaces, this symmetry is broken, and the phenomenon of SHG can occur. This process is termed as surface second-harmonic generation (SSHG). In centro-symmetric crystals such as MgO used in this study, phase matching conditions are not met, and the observed SHG produced is due to SSHG. For example, perturbative and non-perturbative high harmonics have been generated from the surface states of solids which shows the suppression of higher harmonics in bulk material due to lack of phase matching [1]. For an in-depth theoretical and experimental treatment of SHG, we refer the reader to the following works [2-7]. Practical uses of SSHG are widespread, including SHG spectroscopy for imaging to investigate the bio-molecular interactions at interfaces [8], optical imaging [9, 10], characterization of the interface of semiconductors [11], and near-field and far-field optical microscopy of microelectronics structures [12].

---

* Corresponding author, *E-mail* address: mukhtar.hussain@tecnico.ulisboa.pt

Here, we introduce Chromium (Cr) into pure MgO to experimentally investigate the role of doping in the SHG process. Cr is a transition metal, with higher energy occupied electron orbitals than MgO. Doping MgO with Cr will have the effect of adding electrons at normally unoccupied energy levels in the MgO energy states. Chromium atoms have been shown to replace Mg sites during the doping process [13,14] and simulations showed the homogeneous distribution of Cr at the oxide lattice at low doping concentration [14]. The Cr dopants introduce electronic states in the MgO bandgap, which give rise to new optical transitions [15]. MgO doped with Cr, Cr: MgO, shows practical promise in a number of areas. It has been shown to generate less local stress compared to other transition metals when introduced into MgO, making it a good candidate for doping [16]. Cr-doped MgO possesses long-range order and a number of active optical sites, which make it a good candidate for the optical measurements. The manipulation of the optical properties of MgO by introducing Cr dopants has been attributed to the new optical transitions made available from the Cr to the MgO orbitals [13]. The impact of doping concentration and crystal orientation on SHG is both a relevant scientific and practical question. By introducing dopants, we can alter the electronic structure and change the non-linear optical properties of the crystal. This could be used to tailor certain materials for strong field optoelectronics applications, for example a new degree of control in solid-state high harmonics [17].

In this work, we have generated the second harmonic in pure MgO and in Cr: MgO crystals. The impact of crystal orientation and doping concentration on the yield of SHG signal has been investigated. Experimental results have been explained by calculating the electronic structure using density functional theory (DFT) of pure MgO and Cr doped MgO crystals.

In section 2 we describe the experimental approach. The dependence of dopant concentration and crystal orientation on SHG are discussed in 3.1 and 3.2, respectively. We present DFT calculations of the electronic structure of the crystals, and discuss our findings in 3.3. Finally, we give concluding remarks in section 4.

## 2. Experimental setup

We used near-infrared (NIR) laser pulses of 40 fs at 800 nm operating at a repetition rate of 1 kHz, focused on the solid crystals to generate the second harmonic (SH). The schematic of the experimental setup is shown in Fig. 1a. The pulse duration of the driving field was measured by an autocorrelator, as shown in Fig. 1b, and the spectral profile of fundamental field measured by a UV-VIS spectrometer, as shown in Fig. 1c. Femtosecond laser pulses of ~ 20 $\mu$J energy were focused on the 200 $\mu$m thick pure MgO and Cr-doped MgO bulk crystals to ~ 100 $\mu$m diameter by a convex lens of 750 mm focal length. The peak intensity of ~ 1.0 × $10^{13}$ W cm$^{-2}$ is below the damage threshold of MgO. The incident beam was kept normal to the surface plane of the crystal. The crystal was mounted on a three-dimension translation stage as well as on a motorized rotational stage to keep the focal spot fixed at a point in the crystal during the rotation of the target.

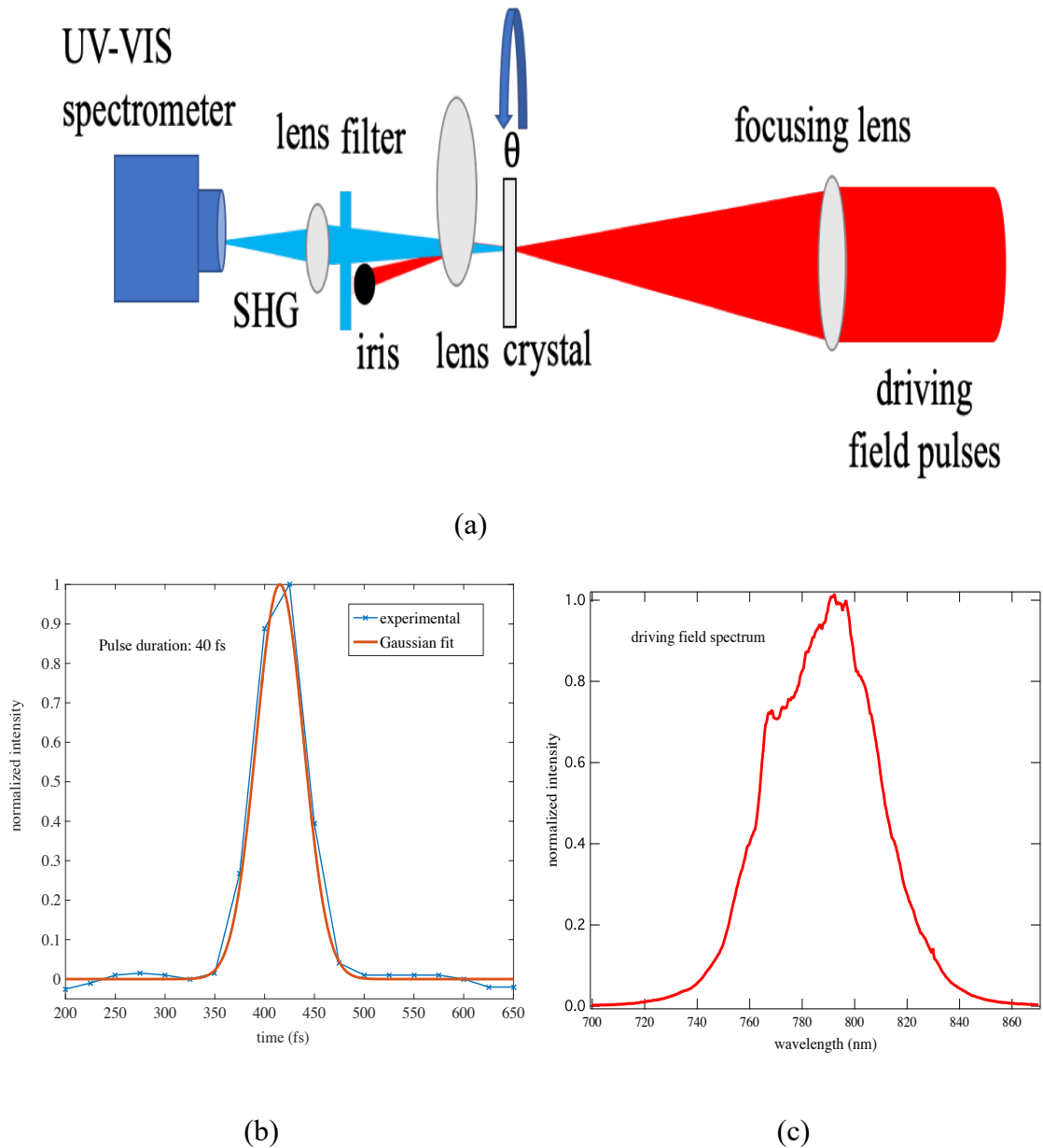

Fig.1: (a) Schematic of second harmonic generation in MgO. The driving laser parameters: 40 fs, central wavelength 800 nm operating at 1 kHz, (b) The pulse duration of driving pulses measured from autocorrelator, (c) Spectrum of the driving pulses.

The MgO crystals are cubic with 001-cut and (100) edge orientation. The pure MgO crystals were fabricated by a three-phase electric arc furnace as a single bulk crystal. The Cr powder is introduced directly into the MgO during the growth process and hence is uniformly distributed throughout the MgO. The fundamental and second harmonic pulses are separated using the chromatic dispersion of the side edge of a convex lens of a 100 mm focal length. The fundamental pulses are blocked by an iris and a short pass filter. The SH pulses are further focused by a 50 mm focal length into the optical fiber, which is connected to a UV-VIS spectrometer. The laser polarisation has been kept fixed while the target rotated about its axis to observe the orientation dependence of SHG in the pure and doped Cr: MgO crystals. The SHG spectrum is measured by a UV-VIS spectrometer. The SHG is first optimized by tuning the laser focus to maximize the SHG yield in pure MgO. These focusing parameters are then kept fixed for all the doped samples.

## 3. Results and Discussion

*3.1 Doping impact on the yield of SHG*

In this section, we investigate the impact of Cr dopant concentration in pure MgO on the SHG efficiency. To validate the wavelength of the SH we present the spectral measurements of the SH at a fixed polarisation of the driving laser pulse, as shown in Fig. 2a. The spectral shape is identical for all dopant concentration, yet the intensity varies significantly. This is consistent with the different overall intensities presented in Fig. 3. About a two-fold higher intensity of SHG is observed for a doping concentration of 740 parts per million (ppm) compared to the pure MgO crystal. Interestingly, as the concentration of Cr is increased further, the SHG intensity monotonically decreases, as shown in Fig. 2b. To observe the global dependence of Cr concentration we present total SHG yield accumulated over all angles for several dopant concentrations. The impact of doping on the yield of SHG is shown in Fig. 2b. The yield is measured as the average over one complete rotation of the crystal. This shows that at the lowest concentration of doping, which is 740 ppm here, the yield is highest. The yield of SHG decreases as the dopant concentration increases further.

To understand the dependence of efficiency on the Cr doping concentration, we now discuss the effect of Cr doping on the electronic, structural and optical properties of MgO, with reference to previous studies. At low dopant concentrations, most of the chromium ions take the position of the Mg ions [18]. Higher dopant concentrations of Cr have been shown to introduce different crystal phases, which may affect the SHG process [15]. Furthermore, it has been reported that by increasing the concentration of Cr doping, the number of atom-sized holes at the surface increases, which could also affect the efficiency of SHG [13]. The atom-sized hole may increase the symmetry breaking and thus enhance the SHG. At increasing levels of dopant concentration, structural changes occur, yet the influence of these new crystal phases on the optical properties is expected to be small. SSHG may not be strongly affected by the overall crystal structure, yet small imperfections at the crystal surface may play a role.

Doping common oxide crystals to alter their optical properties is a widely known and well-developed industrial technique [19, 20]. Here, we look at other works that have independently measured the effect of Cr doping in MgO on the optical properties. We are primarily concerned with our wavelengths of interest (800 nm and 400 nm). Cr: MgO has shown that transmission of optical light decreases overall with increasing doping concentration [21], relevant for both our wavelengths of interest.

The optical properties are a direct consequence of the electronic structure of the material. To clarify how the electronic structure changes in MgO when Cr atoms are introduced, we have conducted DFT simulations of pure and Cr doped MgO. The results of these calculations, along with a discussion of how they explain the experimental results are presented in Section 3.3.

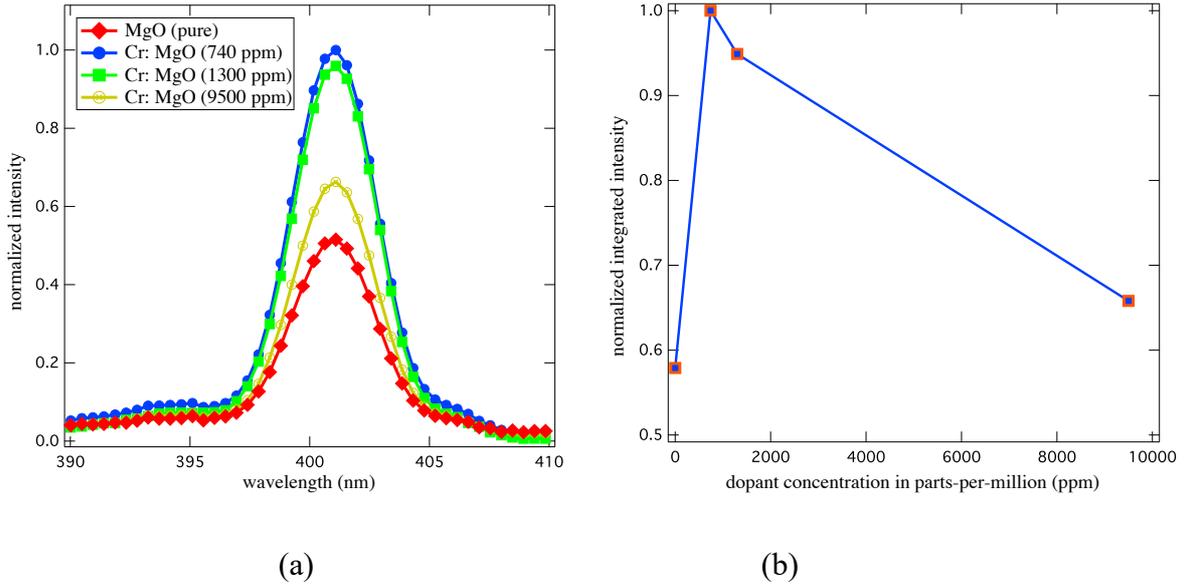

Fig. 2 a) SHG spectrum at a given angle, $\theta$, of rotation of pure MgO and MgO with different dopant concentrations of Cr. Pure MgO crystals (red), 740 ppm of Cr dopant in MgO crystal (blue), 1300 ppm of Cr dopant in MgO crystal (green) and 9500 ppm dopant concentrations of Cr in MgO (yellow), (b) angle averaged SHG intensity as a function of Cr dopant concentration.

## 3.2 Orientation dependence of SHG

At a fixed polarisation of driving laser pulses, the crystals were rotated to observe the crystal orientation dependence of SHG with respect to the electric field. An anisotropic and chaotic dependence of SHG on the crystal orientation is observed in pure MgO as shown in Fig. 3. This polar graph has the linear axis in arbitrary units. The behaviour shown here underlines the complex dynamics of SHG in pure MgO and is attributed to the sensitivity of surface structural effects [22-24]. The exact shape of the orientation dependence of SHG may vary depending on crystal surface features or quality, as SH is generated due to the roughness of crystal surface. Conversely, isotropic emission is observed for all crystals containing Cr dopant, as shown in Fig. 3.

Furthermore, the Cr-doped MgO samples show a higher SH efficiency for all dopant concentrations. The angular anisotropy of the second harmonic emission relative to the crystal orientation present in pure MgO is lost when dopants are introduced. We have found that the exact angular dependence pattern of pure MgO to be sample dependent. Moreover, SHG was not observed from MgO samples with two polished surfaces, which shows that the disordered surface morphology is the source of SHG in this study. However, the overall efficiency of SHG is dependent on the doping concentration. This has been discussed in section 3.1.

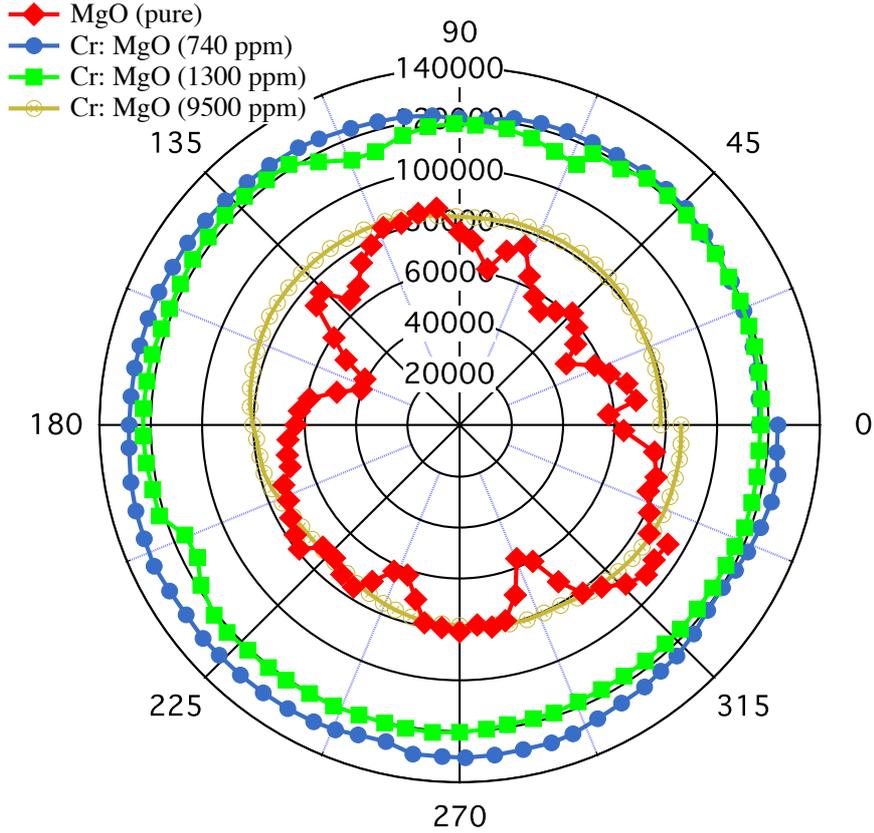

Fig. 3 Orientation dependence of SHG in undoped and Cr-doped MgO crystal at fixed driving laser polarisation and rotation of crystal about its center. Pure MgO crystal (red), 740 ppm of dopant (blue), 1300 ppm of dopant (green) and 9500 ppm of dopant (yellow).

### 3.3 Electronic Structure Calculations from Density Functional Theory

The SHG process has an inherent dependence on the allowed energy states and their corresponding occupation in the crystal. To better understand the role of the Cr doping, we have performed density functional theory (DFT) calculations of the pure and doped crystals (shown in Fig. 4(a) and (b), respectively). The DFT calculations were performed using the VASP package [25-27]. Projected augmented wave (PAW) pseudo-potentials [28] and the generalised gradient approximation (GGA) for the exchange-correlation functional were used in combination with 300 bands and a $12 \times 12 \times 12$ automatically generated k-point mesh. The recently measured valence band structure of MgO agrees with the our DFT calculations [29]. The single difference in the calculations of the pure MgO and doped Cr: MgO cases here was the inclusion of a Cr atom at the central site normally occupied by Mg. The cell size was 64 atoms in total, therefore introducing a Cr atom at the centre of this cell would correspond to a ~ 1.6% dopant concentration. This is higher than our maximum dopant concentration of ~1% in the experiment. However, the increase in cell size necessary for calculations with lower dopant concentrations becomes computationally prohibitive. Nonetheless, the physical effect of the dopant on the electronic structure remains qualitatively valid. A further increase in dopant concentration would increase the occupation of higher energy states, and ultimately converge to the electronic structure of metallic Cr.

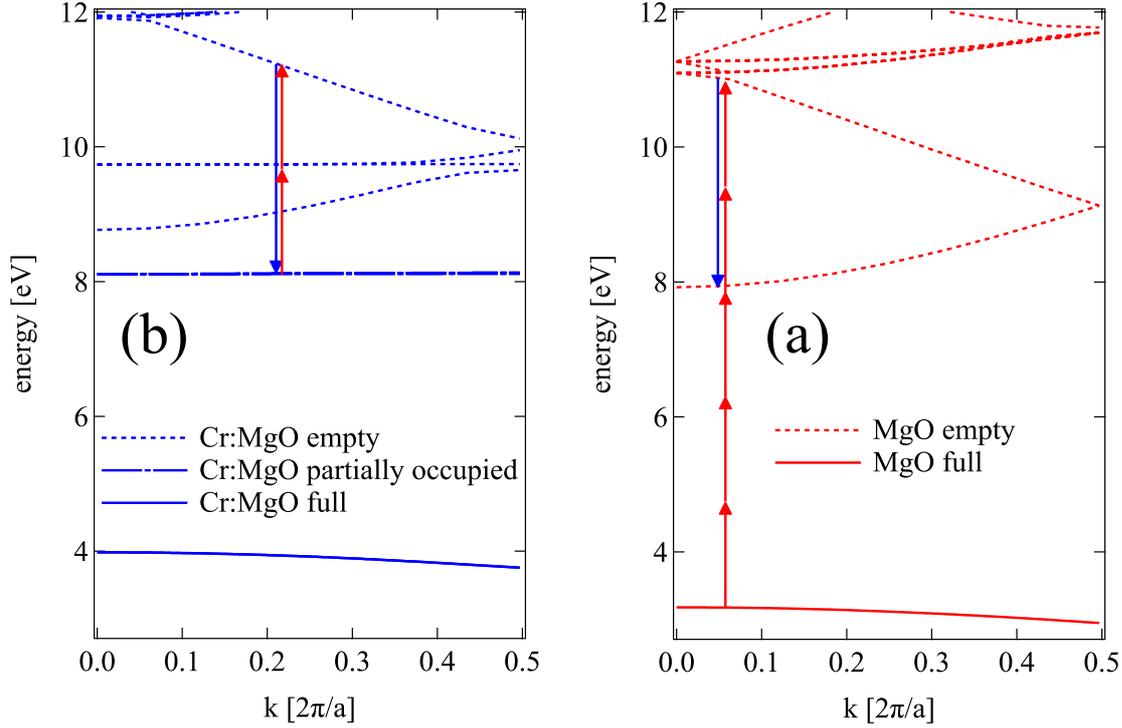

Fig. 4 The band dispersion plots using DFT, calculated in the Γ-X direction for (a) pure, and (b) chromium doped, MgO. The lattice constant used was $a$=4.2 Å. The red arrows indicate the driving laser photon energy, and the blue arrows show the photon energy of the second harmonic. The occupation of the various bands is indicated in the respective legends.

We present calculations of the band dispersion in the Γ-X direction of pure and Cr doped MgO in Fig.4a and b, respectively. In pure MgO (Fig.4a), we note an optical band gap of bulk MgO of around 4.7 eV. This lies in clear contrast to the accepted band-gap of around 7.8 eV [30]. The discrepancy between the accepted band-gap and the value reported here is due to the choice of xc functional used. Here we use the GGA to the xc functional, which is known to underestimate the band-gap. In fact, calculations similar to those presented here have yielded similar bulk band-gaps from 4.6 to 5 eV for bulk MgO [30,31]. It should be noted that surface band gaps of MgO have been shown elsewhere to be in the 5 eV range. Coincidentally, the bulk DFT band-gaps reported here are closer to the measured surface MgO band-gaps, where the SHG process reported here largely occurs.

It has been shown elsewhere that band-gaps of MgO can be well reproduced with other, more computationally expensive methods [30]. Calculations of the band-structure of MgO have been shown to match experimental values when more complex approaches (using hybrid Hartree-Fock methods) to the xc functional have been employed [31]. However, these approaches have the downside of often been computationally more intensive. Since our cell size is already at the limit of our computational resources, we restrict our calculations to the more efficient, but less accurate GGA approach. Despite the expected underestimation of the optical band-gap in bulk MgO reported here, we maintain that this difference does not affect our conclusions: the reduced the band-gap by introducing Cr dopants increases the photon absorption process, thereby increasing the SHG efficiency.

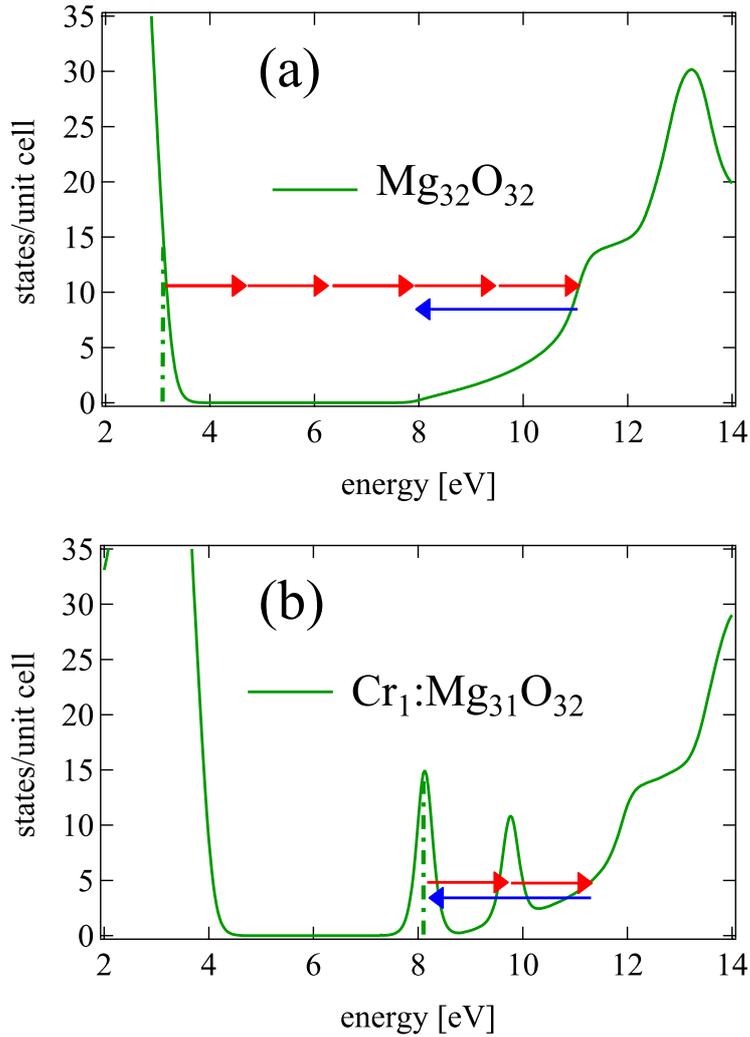

Fig. 5: The density of states of states of MgO (a) and Cr: MgO (b) calculated with DFT. The red arrows indicate the energy of a photon of the fundamental IR driving laser field. The blue arrows indicate the energy of a second harmonic photon.

In Fig. 5 the highest occupied energy level (Fermi energy) is indicated by the vertical dash-dot line around 3 eV in 5(a). There are no available energy levels for blue (SHG) photon transitions to occur around the Fermi energy for pure MgO. Therefore, a multi-photon absorption process must occur to drive electrons to the higher energy states (red arrows in 5(a)). Conversely, the Cr: MgO has higher lying occupied (defect) states, due to the Cr doping (sharp peak around 8 eV). This higher lying occupied (defect) state allows for direct driving and recombination of electrons, without the need for prior multi-photon absorption processes to occur. This increase efficiency by allowing a more probable two-photon process to excite the electrons, as shown in Fig. 2. The dependence of SHG efficiency on the direction of the driving field is minimal, as the highest occupied state (dashed line in 5(b) is a narrow, atomic like line that varies little in energy with respect to crystal orientation (as shown in Fig. 3). This lack of angular dependence on the initial and final state in the electronic structure translates to a lack of observed angular dependence with Cr: MgO, as shown in Fig. 3.

Hence, by manipulating the electronic structure, we can tailor the efficiency and polarisation dependence of non-linear processes in crystals. The efficiency, however, can be limited by the changing optical properties, as the dopant concentration is increased. This shows a delicate balance of the element and concentration of the dopant is essential for improving solid-state SHG in terms of yield.

## 4 Conclusion

We have generated the second harmonic in pure and Cr-doped MgO with a NIR driving field. On rotating the crystal relative to the driving field polarisation, an anisotropic and chaotic behavior of SHG efficiency in pure MgO is observed. This is attributed to the complex interplay between surface and non-linear effects for SHG in MgO. Conversely, the SHG signal was isotropic and symmetrical in Cr doped MgO crystals. This isotropic emission of SHG is attributed to the minimal angular dependence of the Cr valence band, which we identify as the most probable initial energy state of the multi-photon IR absorption process through DFT calculations. The effect of doping concentration on the yield of SHG signal has been investigated, showing that the lowest (740 ppm) concentration of Cr gives a two-fold increase in SHG efficiency relative to pure MgO. At higher doping concentrations the SHG yield is reduced. This reduction in SHG yield at higher dopant concentrations is attributed to the changing linear optical properties of doped MgO. In particular, increased re-absorption of the SH increases with Cr dopant concentration, and therefore limits the overall achievable SHG efficiency.

This work has shown that, by introducing dopants into simple crystals, the electronic structure can be shaped to tailor the non-linear optical response in terms of efficiency and angular dependence. We show that Cr can increase the SHG efficiency and mitigate angular polarisation dependence when introduced to MgO, and that efficiency favors dopant concentration on the order of 740 ppm. This study paves the way for the merging of bandgap engineering and solid-state harmonic generation to tailor frequency up-conversion processes. Perhaps of equal importance, is the prospect of experimentally testing theories of the solid-state high harmonic generation (HHG) mechanism with tailored electronic structures, with a final view towards all optical mapping the band structure of solids.

**Acknowledgments**

This work was partially supported by the Fundação para a Ciência e Tecnologia (FCT) under the grant number PD/BD/135224/2017 in the framework of the Advanced Program in Plasma Science and Engineering (APPLAuSE), and the PETACom FET Open H2020, support from the French ministry of research through the ANR grants 2016 "HELLIX", 2017 "PACHA", the DGA RAPID grant "SWIM" and the LABEX "PALM" (ANR-10-LABX-0039-PALM) through the grants "Plasmon-X", "STAMPS" and "HILAC". We acknowledge the financial support from the French ASTRE program through the "NanoLight" grant. We are thankful to Dr. Viktoria Nefedova and Dr. David Gauthier for fruitful discussion.